\documentclass[twocolumn]{svjour3}         
\smartqed  
\usepackage{graphicx}
\usepackage{mathptmx}      
 \journalname{Astrophysics and Space Science}
\begin{document}

\title{A large  CO and HCN line survey of Luminous Infrared Galaxies}
\titlerunning{CO and HCN line survey of LIRGs}   
\author{Padelis P. Papadopoulos \and Thomas R. Greve \and Paul van der Werf 
\and Stefanie M\"uehle \and Kate Isaak \and Yu Gao }


\institute{Padelis P. Papadopoulos \at
  Institut f\"ur Astronomie, ETH Zurich, Zurich, Switzerland\\
  Tel.: +41 (0)44 633 3826  \\
  Fax: +41 (0)44 633 1238  \\
  \email{papadop@phys.ethz.ch}    \\
  Thomas R. Greve \at California Institute of Technology, Pasadena, CA 91125, USA\\
  Paul van der Werf \at Sterrewacht Leiden, 2300 RA Leiden, The Netherlands\\
  Stefanie M\"uehle \at Department of Astronomy,
  Univ. of Toronto, 60 St. George Street, Toronto, ON M5S 3H8, Canada\\
  Kate Isaak \at Physics and Astronomy, Univ. of Wales, Cardiff CF24 3YB,~UK\\
  Yu Gao \at Purple Mountain Observatory, Chinese Academy of Sciences, Nanjing
  210008, People's Republic of China
  }

\date{Received: date / Accepted: date}

\maketitle

\begin{abstract}
  A  large CO,  HCN multi-transition  survey of  30  Luminous Infrared
  Galaxies  ($\rm  L_{IR}>10^{11}\,L_{\odot}$)  is nearing  completion
  with the James Clerk Maxwell Telescope (JCMT) on Mauna Kea (Hawaii),
  and  the IRAM  30-meter telescope  at  Pico Veleta  (Spain). The  CO
  J=1--0, 2--1, 3--2, 4--3,6--5,  $ ^{13}$CO J=2--1, HCN J=1--0, 3--2,
  4--3 observations,  resulting from $\sim  250$ hours of  JCMT, $\sim
  100$  hours of  30-m observing  time and  data from  the literature
  constitute {\it  the largest extragalactic molecular  line survey to
  date},  and can  be  used to  address  a wide  range  of issues  and
  eventually allow  the construction of reliable  Spectral Line Energy
  Distributions  (SLEDs) for  the molecular  gas in  local starbursts.
  First  results  suggest  that:   a)  HCN  and  HCO$^+$  J=1--0  line
  luminosities  can be  poor mass  estimators of  dense  molecular gas
  ($\rm  n\geq 10^4\,cm^{-3}$)  unless their  excitation  is accounted
  for, b) CO  cooling of such gas in ULIRGs may  be comparable to that
  of the  CII line at $\rm 158\,\mu  m$, and c) low  excitation of the
  {\it  global}  molecular  gas  reservoir remains  possible  in  such
  systems.  In such  cases the expected low CO  $\rm J+1\rightarrow J$
  line  luminosities for $\rm  J+1\geq 4$  can lead  to a  strong bias
  against their detection from ULIRGs at high redshifts.

\end{abstract}

\section{Introduction}
\label{intro}
\vspace*{-0.20cm}  The   importance  of  Luminous   Infrared  Galaxies
(hereafter  LIRGs)  as  the  sites  of the  most  extreme  local  star
formation  events  (\cite{Sand04})   makes  them  the  best  available
templates for similar events  in the distant Universe.  Moreover their
compact  CO-emitting  regions   (\cite{DS98})  makes  them  ideal  for
multi-line studies of their global  molecular gas reservoir up to very
high frequencies (where the high-excitation molecular lines lie) since
the resulting narrow beams of today's single dish radio telescopes can
still  measure their  total line  fluxes with  single  pointings.  Our
sensitive  CO J=2--1,  3--2, 4--3,  6--5,  $ ^{13}$CO  J=2--1 and  HCN
J=3--2  and 4--3 line  observations with  the JCMT  and the  IRAM 30-m
telescope  for $\sim  30$ LIRGs,  combined with  available CO  and HCN
J=1--0  data  from  the  literature,  will  allow  for  excellent  new
constraints  on the state  of their  global molecular  gas reservoirs.
More  specifically the  high-J CO  and the  three HCN  transitions can
probe the excitation and mass of the dense ($\rm n\geq 10^4\,cm^{-3}$)
gas,  considered  as  the  immediate  fuel of  their  prodigious  star
formation  (\cite{Sol92},  \cite{Gao04}).  The  completion  of our  CO
J=4--3  and J=6--5  observations and  follow-up observations  of still
higher  rotational  CO,  $  ^{13}$CO  and HCN  transitions  with  {\it
Herschel} will ultimately allow the  deduction of robust SLEDs for the
star-forming molecular gas of LIRGs  in the local Universe.  These can
then   be   used   to   uncover   any   universal   aspects   of   the
star-formation/molecular gas  interplay in galaxies and  find the best
rest-frame  mm/sub-mm lines  for  detecting and  imaging the  numerous
LIRGs that  will be found at  high redshifts in  future deep mm/sub-mm
continuum surveys (\cite{Hughes04}).

\section{HCN versus HCO$^+$ lines as dense gas mass tracers}
\label{sec:1}

\begin{figure*}
\begin{center}
\includegraphics[width=0.44\textwidth]{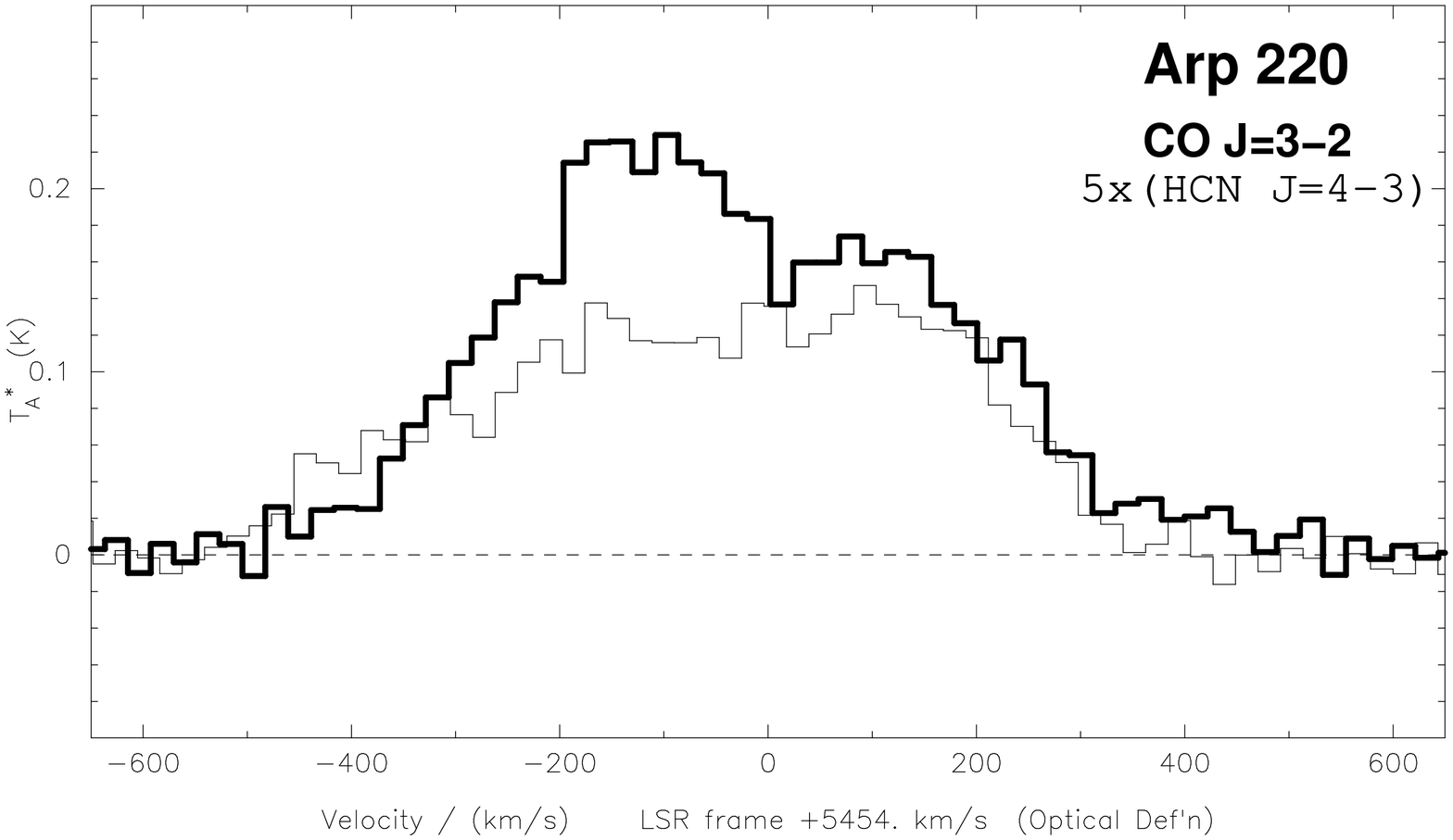}
\includegraphics[width=0.44\textwidth]{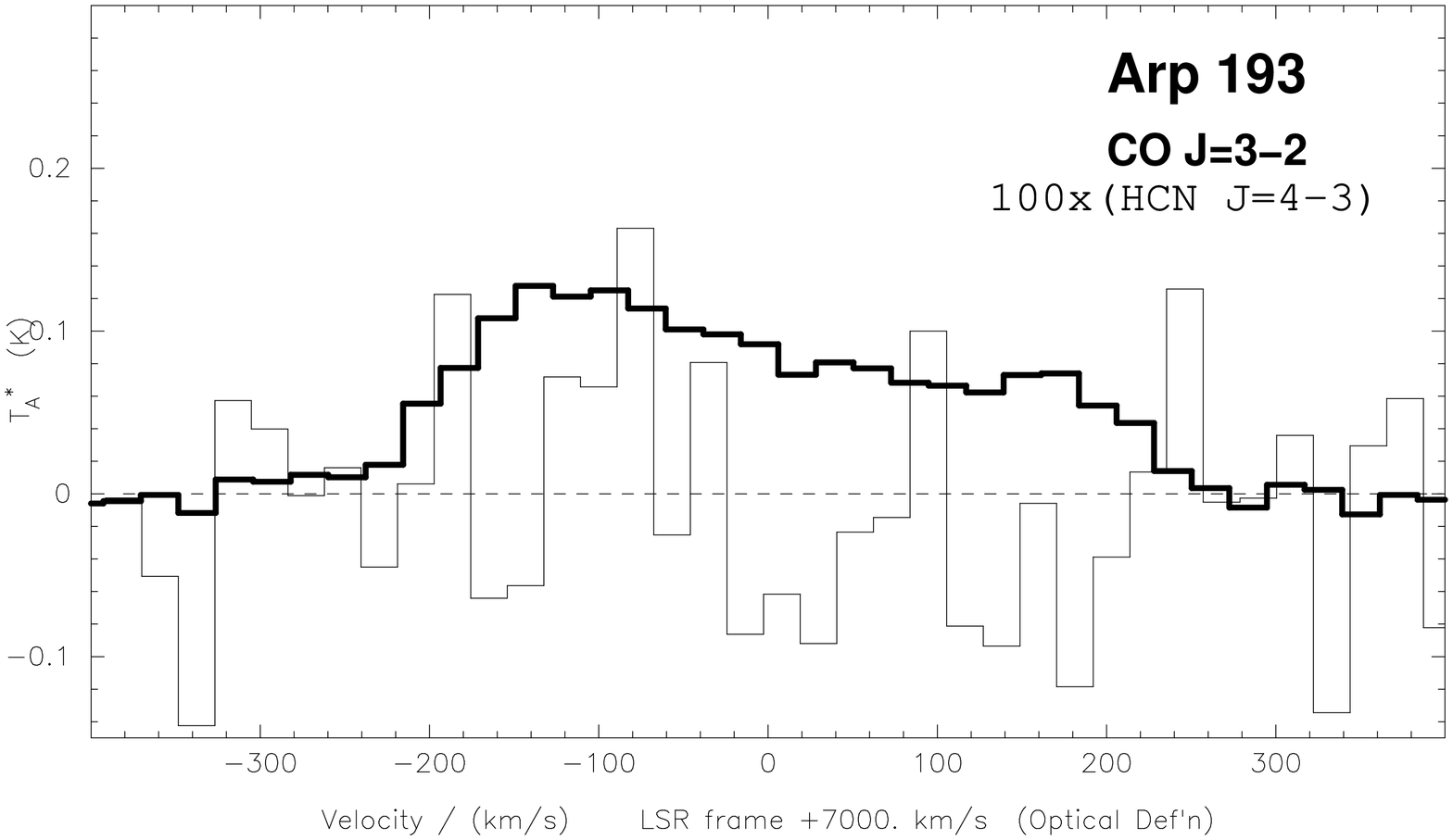}\\
\end{center}
\caption{HCN J=4--3 (scaled by the numbers in the upper right of each panel) and
CO J=3--2 spectra of four LIRGs. The HCN (4--3)/(1--0) brightness temperature ratio
$\rm r_{43}(HCN)$ is found to vary from $\rm r_{43}(HCN)\leq 0.1$ (Arp\,193) to
 $\rm r_{43}(HCN)\sim 0.8$ (Arp\, 220).}
\label{fig:1}       
\end{figure*}

The rotational transitions of the HCN and HCO$^+$ molecules with their
high dipole moments are currently the best (i.e.  most luminous) dense
gas  tracers in  LIRGs. An  HCN J=1--0  line survey  of  such galaxies
(\cite{Gao04}) has even revealed a potentially constant star formation
efficiency ($\propto \rm  L_{IR}/M_{dense}(H_2)$) for molecular gas at
densities $\rm n\geq  10^4\, cm^{-3}$, attributed to such  a gas phase
being  the   direct  fuel  of  star  formation   in  molecular  clouds
(\cite{Wu05}).  Recent studies have  inserted some doubt as to whether
HCN lines are good tracers of  such a gas phase and suggested those of
HCO$^+$ instead (\cite{Carp06}).  Our  survey reveals a large range of
HCN line  excitation in  LIRGs with otherwise  similar IR, CO  and HCN
J=1--0 luminosities (Figure 1).  In  the case of Arp\,193 the observed
very low  HCN (4--3)/(1--0) line ratio  (its HCN 4--3  is $>$100 times
weaker than its CO J=3--2 line!)  {\it is compatible with the complete
absence of  a massive  and dense molecular  gas phase.}  On  the other
hand  for the  two archetypal  ULIRGs Arp\,220  and NGC\,6240  the HCN
ratios  are  significantly  larger  than the  corresponding  ones  (in
rotational level)  of HCO$^+$  (Table 1).  This,  along with  the fact
that HCO$^+$  rotational transitions also have $\sim  5-7$ times lower
critical densities  than those  of HCN, signifies  that the  latter is
tracing  a denser  gas  phase and  may  thus may  remain  as the  most
suitable  gas mass  tracer  in  galaxies once  the  excitation of  its
rotational lines is properly accounted for.

\begin{table}
\caption{HCN vs HCO$^+$ line ratios in Arp\,220 and NGC\,6240}
\label{tab:1}       
\begin{tabular}{lllll}
\hline\noalign{\smallskip}
Galaxy & $\rm HCN\frac{(4-3)}{(1-0)}$ &$\rm HCN\frac{(3-2)}{(1-0)}$ & 
$\rm HCO^+\frac{(4-3)}{(1-0)}$ & $\rm HCO^+\frac{(3-2)}{(1-0)}$  \\
\noalign{\smallskip}\hline\noalign{\smallskip}
Arp\,220  & $0.8\pm 0.2$ & $1.0\pm0.3 $  & $0.33\pm0.10$ & $0.27\pm 0.10$ \\
NGC\,6240 & $0.6\pm 0.2$ & $1.0\pm 0.3$  & $0.21\pm0.06$ & $0.24\pm0.08$ \\
\noalign{\smallskip}\hline
\end{tabular}
\end{table}

\section{The dense molecular gas in Mrk\,231}
\label{sec:2}

\begin{figure}
\includegraphics[width=0.49\textwidth]{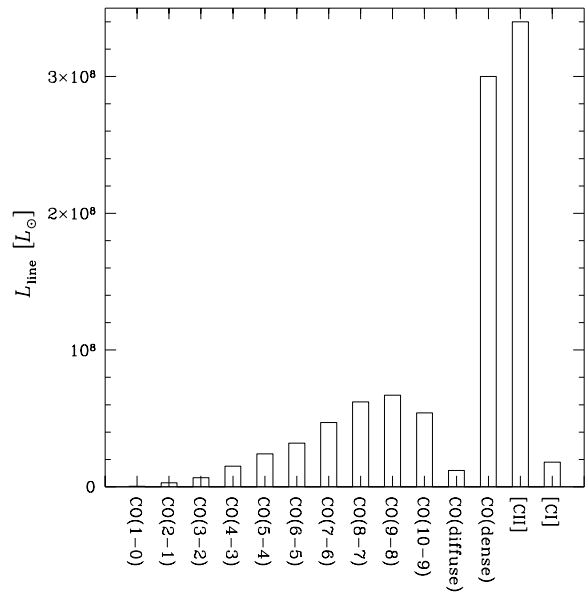}
\caption{CO, CI, CII cooling contributions in the ULIRG/QSO Mrk\,231}
\end{figure}

The prototypical ULIRG/QSO Mrk\,231 is  the first object in our sample
for  which the  full  set  of line  observations  has been  completed,
allowing for  a unique view  into the global excitation  conditions of
its  molecular gas.   Analysis of  the  relative strengths  of the  CO
J=1--0,  2--1, 3--2,  4--3, 6--5  and HCN  J=1--0, 4--3  lines  (and $
^{13}$CO J=2--1 from the  literature,\cite{Glen01}) finds the low-J CO
lines dominated by diffuse  ($\rm n\sim few\times 10^2\,cm^{-3}$), and
warm  ($\rm  T_k\sim 90-100\,K$)  gas,  while  the high-excitation  CO
J=6--5,  4--3  and the  HCN  transitions  trace  a denser  ($\rm  \sim
(1-3)\times  10^4\,cm^{-3}$) phase  with  somewhat lower  temperatures
($\rm T_k\sim  50-70\,K$).  The  latter dominates the  total molecular
gas mass in Mrk\,231, and its total CO line cooling is similar to that
of the  CII line at  $158\,\mu m $  (Figure 2).  {\it This  suggests a
different thermal  balance to  that of lower  IR-luminosity galaxies,}
and may explain the relative weakness  of the CII cooling line in such
systems as a result of the dominance of dense Photon Dominated Regions
(PDRs) for most of their molecular gas (\cite{Kau99}).

\subsection{A comparison with high-z starbursts}
\label{sec:2}

\begin{figure}
\includegraphics[width=0.47\textwidth]{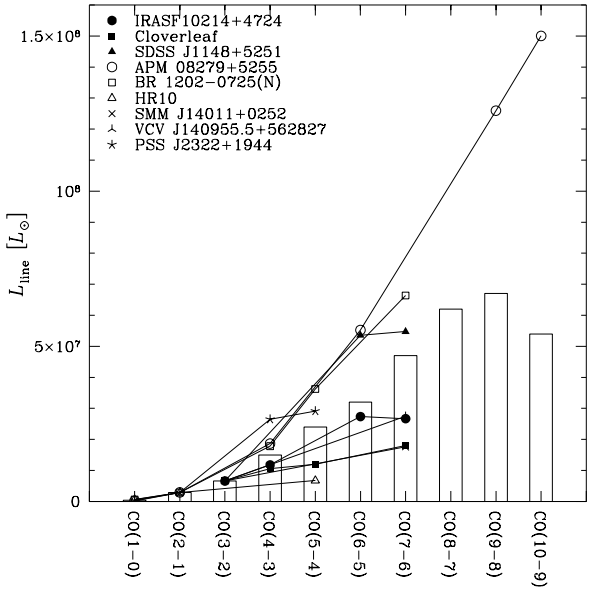}
\caption{CO  line   luminosities  of  various   high-redshift  objects
  compared to the Mrk\,231 CO dense gas SLED template (bars).
  Luminosities are normalized to those of Mrk\,231  at either CO
  J=3--2 or  J=2--1.  Line  fluxes for the  various objects  are taken
  from the literature (\cite{Sol05}, \cite{Riec06}).}
\label{fig:2}       
\end{figure}

The  high-J CO and  the HCN  lines detected  in Mrk\,231  provide good
constraints on  radiative transfer models  which then allow a  good CO
SLED to be constructed.  Its comparison with objects at high redshifts
shows that despite the high  excitation of its dense molecular gas its
CO SLED stands  as rather average (Figure 3).  However, strong lensing
(affecting several  objects in Figure  3) can selectively  amplify the
highly excited  CO line emission  of compact starburst regions  at the
expence  of  a  much  more extended  low-excitation  non-star  forming
molecular gas  phase (e.g.  \cite{Weiss04}).   Furthermore, the high-J
CO  observations usually  conducted after  spectroscopic  redshifts of
high-z objects  become available  may select mostly  from a  subset of
starbursts where the  (star-forming)/(non star-forming) H$_2$ gas mass
ratio (expected to  vary strongly in the several  merger systems among
LIRGs) is  particularly large.  The limited sensitivity  of today's cm
and mm/sub-mm single dish  telescopes and interferometers will further
accentuate the  aforementioned biases, which may thus  {\it produce CO
SLEDs at high  redshifts which are biased towards  the high excitation
regime}, and are thus unrepresentative of the full excitation range in
star-forming galaxies.

In that respect  CO (and HCN) SLEDs of  local ULIRGs, established from
observationally well-sampled J-ladders  of rotational transitions, are
indispensible as benchmarks unaffected by the aforementioned biases.

\section{A ``cold'' and a ``warm'' ULIRG}

\begin{table}
\caption{CO  line ratios of a ``warm'' and a ``cold'' ULIRG}
\label{tab:1}       
\begin{tabular}{lllll}
\hline\noalign{\smallskip}
Galaxy & $\rm CO\frac{(2-1)}{(1-0)}$ &$\rm CO\frac{(3-2)}{(1-0)}$ & 
$\rm CO\frac{(4-3)}{(1-0)}$ & $\rm \frac{^{12}CO}{^{13}CO}$  \\
\noalign{\smallskip}\hline\noalign{\smallskip}
17208-0014  & $1.02\pm 0.22$ & $0.72\pm 0.17 $  & $0.74\pm 0.23 $ & $\geq 35$ \\
05189-2524  & $0.32\pm 0.08$ & $0.24\pm 0.07$  & $<0.32$ & $6\pm 2$  \\
\noalign{\smallskip}\hline
\end{tabular}
\end{table}

Molecular gas excitation  at levels far below those  found in Mrk\,231
or other such galaxies used  as templates for high redshift starbursts
(e.g.  Arp\,220),  have been uncovered by our  ongoing survey, further
underlying the need  for local CO SLEDs to be  established for a large
sample of local LIRGs.

The   two    ULIRGs   ($\rm   \rm    L_{IR}>   10^{12}\,   L_{\odot}$)
IRAS\,17208-0014  and IRAS\,05189-2524  have been  found to  have very
different CO  line excitation (Figure 4) despite  similar IR continuum
and  CO   J=1--0  line   luminosities  and  similar   $\rm  S_{60\,\mu
m}/S_{100\,\mu  m}$  ratios. For  IRAS\,17208-0014  the CO  brightness
temperature ratios  measured are typical for  the high-excitation star
forming molecular gas found in many such galaxies, but those found for
the  ``cold''   IRAS\,05189-2524  are  surprisingly   typical  of  the
quiescent  molecular clouds  found in  the Milky  Way disk  (Table 2).
One-phase radiative transfer models  using the Large Velocity Gradient
(LVG)   approximation  and   constrained   by  the   line  ratios   of
IRAS\,17208-0014   are  compatible  with   densities  of   $\rm  n\sim
10^3\,cm^{-3}$ and mostly warm  $\rm T_k\sim (70-100)\,K$.  The high $
^{12}$CO/$  ^{13}$CO  J=2--1 ratio  measured  for  this merger  system
(Table 2)  is responsible for the  low CO J=1--0  optical depths ($\rm
\tau  _{10}\leq  0.5 $)  of  this  gas phase,  a  result  of its  high
temperatures  and  large  average  velocity gradients  $\rm  dV/dR\sim
30\,km\,s^{-1}\,pc^{-1}$ (for an abundance of $\rm [CO/H_2]=10^{-4}$).
The      latter     corresponds      to      a     parameter      $\rm
K_{vir}=\frac{(dV/dR)_{obs}}{(dV/dR)_{virial}}\sim 30-50 $, typical of
an unbound gas phase (\cite{Pap99}).  It is also worth noting that the
high (4--3)/(1--0)  CO ratio  (comparable to that  of CO(3--2)/(1--0))
cannot be accounted by a single gas phase model and probably signifies
the emergence of  a second warmer and denser  phase.  In several LIRGs
(e.g.   Mrk\,231) such  a phase  has been  found to  dominate  CO $\rm
J+1\rightarrow  J$,  $\rm  J+1\geq   4$  and  HCN  line  emission  and
containing most  of their molecular gas  mass, and is  most likely the
direct fuel of their prodigious star formation.

\begin{figure*}
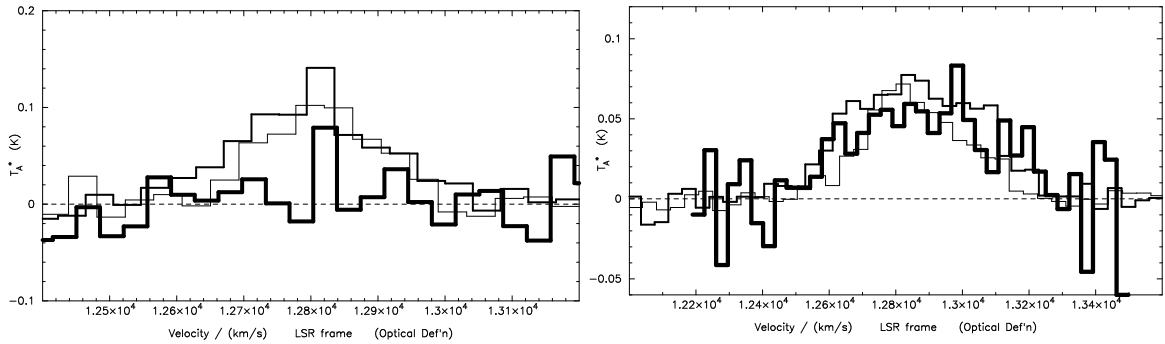

\begin{center}
\includegraphics[angle=-90, width=0.44\textwidth]{./IRAS05189_CO4321.ps}
\includegraphics[angle=-90, width=0.44\textwidth]{./IRAS17208_CO4321.ps}
\end{center}
\caption{CO  J=4--3,  3--2  and  2--1  spectral  lines  (plotted  with
declining  line  thickness) for  the  ``cold'' IRAS\,05189-2524  (left
panel) and the ``warm'' IRAS\,17208-0014 (right panel) galaxies.}
\label{fig:1}       
\end{figure*}

The CO  emission in IRAS\,05189-2524  on the other hand  is compatible
with gas  of $\rm n\sim  100\,cm^{-3}$, $\rm T_k\sim 15\,K$,  and $\rm
K_{vir}\sim 1$,  typical of the quiescent  and mostly self-gravitating
molecular  clouds  found  in  the  Galactic disk  immersed  in  a  FUV
radiation  field of  $\rm G_{\circ}\sim  1$ (Habing  units).   In such
conditions high-J  CO lines  are expected to  be very faint  with $\rm
CO(J+1-J)/(1-0)\leq  0.01$ for  $\rm J+1\geq  4$ for  the bulk  of the
molecular gas.  This is much lower than that expected from CO SLEDs of
other ULIRGs such as Mrk\,231 or Arp\,220 or indeed most high redshift
objects  (\cite{Riec06}), with  only the  extremely red  object HR\,10
coming close (Figure 3).

  Thus starburst galaxies at high  resdshifts with similar CO SLEDs to
that  of  IRAS\,05189-2524 {\it  would  remain  undetected by  typical
high-J  CO observations  with  today's radio  telescopes,} even  after
scaling  their CO  line  fluxes by  their  $\sim 10$  times higher  IR
luminosities.  One  of the  goals of the  CO and  HCN multi-transition
survey of  LIRGs is  to find how  often such  low-excitation molecular
SLEDs occur in starbursts, and seek out correlations with other galaxy
characteristics   (e.g.    merger    status,   starburst   age,   dust
temperatures).  This  may in turn  shed some light into  how otherwise
vigorously   star-forming  galaxies  can   harbor  large   amounts  of
low-excitation molecular gas.

\section{Conclusions}

We   report  preliminary  results   from  our   ongoing  CO   and  HCN
multi-transition of  30 local LIRGs, the largest  such  survey to date. 
These can be summarized as follows,

\begin{itemize}
\item A large range of HCN line excitation precludes the simple use of
      the HCN  or HCO$^+$ J=1--0 line  luminosity as a  dense gas mass
      tracer.

\item In the  two cases of Arp\,220 and NGC\,6240,  the HCN versus the
      HCO$^+$ line excitation suggests  the former tracing the densest
      gas phase.

\item In the ULIRG/QSO Mrk\,231 the luminous CO J=4--3 and J=6--5 line
      emission  emanates  from a  different  gas  phase  than the  one
      dominating the low-J CO transitions. In that phase total CO line
      cooling is comparable  to that of the CII  line at $\rm 158\,\mu
      m$.  If  confirmed  for  more  such  galaxies  this  raises  the
      possibility of  a very different thermal balance  than that seen
      in  lower  IR-luminosity systems  where  the  CII  line is  the
      dominant coolant.

\item  The  discovery   of  very  low  CO  excitation   in  the  ULIRG
      IRAS\,05189-2524  raises the possibility  of a  large excitation
      bias  against detecting  similar objects  at high  redshifts via
      their high-J CO line emission.

\end{itemize}

\begin{acknowledgements}
We would like  to thank the telescope operators and  staff of the JCMT
and  IRAM 30-m  telescope for  their assistance  during  the extensive
observations that  made this project possible.  Special  thanks to Jim
Hoge, Per Friberg  and Iain Coulson at the JCMT  for expert advice and
assistance with the observations during many years.
\end{acknowledgements}

\end{document}